\documentclass{article}

\usepackage{amsmath,amssymb,amsfonts,mathrsfs,amsthm,mathdots}
\usepackage{physics}
\usepackage{tensor}
\usepackage{multirow}
\usepackage{graphicx}
\usepackage{caption} 
\usepackage{subcaption} 

\usepackage{biblatex}
\usepackage{lscape}
\usepackage{geometry}
\usepackage{appendix} 
\usepackage{authblk}

\newtheorem{example}{Example}

\title{Exact lambdavacuum solutions in higher dimensions}
\date{March 22, 2026}

\author[1]{I. A. Sarmiento-Alvarado\thanks{ignacio.sarmiento@cinvestav.mx}}
\author[2]{P. Wiederhold\thanks{petra.wiederhold@cinvestav.mx}}
\author[1]{T. Matos\thanks{tonatiuh.matos@cinvestav.mx}}
\affil[1]{Departamento de F\'{\i}sica, Centro de Investigaci\'on y de Estudios Avanzados del Instituto Polit\'ecnico Nacional, Avenida Instituto Polit\'ecnico Nacional 2508, San Pedro Zacatenco, M\'exico 07360, CDMX.}
\affil[2]{Departamento de Control Autom\'atico, Centro de Investigaci\'on y de Estudios Avanzados del Instituto Polit\'ecnico Nacional, Avenida Instituto Polit\'ecnico Nacional 2508, San Pedro Zacatenco, M\'exico 07360, CDMX.}

\begin{document}

\maketitle

\begin{abstract}
    In this work, we obtain exact solutions to the $(n+2)$-dimensional Einstein Field Equations with a non-zero cosmological constant for $n > 1$.
    These solutions depend on a set $\{ A_a, a=1,2,\ldots , m \}$ of pairwise commuting constant matrices in $\mathfrak{sl} ( n, \mathbb{R} )$ and on a constant matrix $g_0$ in $\mathcal{I} (\{ A_a, a=1,\ldots , m \})$, determined in previous work.
    Different choices of $\{ A_a, a=1,\ldots , m \}$ and $g_0$ correspond to different solutions.
    As examples, we show how to obtain the de Sitter metric, the Anti-de Sitter metric, the Birmingham metric, the Nariai metric and the Anti-Nariai metric in higher dimensions.
    The generalized Nariai and Anti-Nariai solutions are direct topological products of $AdS_{\frac{n}{2} + 1} \times H^{\frac{n}{2} + 1}$, $dS_{\frac{n}{2} + 1} \times S^{\frac{n}{2} + 1}$, $AdS_2 \times H^n$, $AdS_n \times H^2$, $dS_2 \times S^n$ and $dS_n \times S^2$.
    In addition, we study a solution in the context of cosmology.
\end{abstract}

\section{Introduction}

In 1917, Einstein introduced the cosmological constant to describe a static universe \cite{1917Einstein}. However, the first solution to Einstein's field equations (EFE) with a cosmological constant was found by de Sitter in 1917 \cite{10.1093/mnras/78.1.3}. These solutions give rise to a new paradigm in which they seek a static universe, just as our universe was believed to be at that time.
In 1918, Kottler found the extended Schwarzschild metric, which includes the cosmological constant \cite{https://doi.org/10.1002/andp.19183611402}. This solution showed that if a star has a cosmological constant inside it, it could have very different properties because of that constant.
In this line of research, Carter discovered the extended Kerr metric with cosmological constant in 1973 \cite{Carter:1973rla},
later, the external solution due to Reissner-Nordström with cosmological constant was obtained by Kramer et al. in 1980 \cite{exact1980}.
Today we know several extended solutions with cosmological constant, some examples are Xu et al. found the extended Florida's solution with a cosmological constant in 1987 \cite{Chongming1987}
and in 1989, Krori et al. generalized the Kottler and Florida solutions to higher dimensions \cite{Krori1989}.

Currently, the cosmological constant is very important in cosmology because it is a strong candidate for dark energy, which makes the universe expand at an accelerating rate.
Exact solutions to EFE with a non-zero cosmological constant provide not only precise theoretical models for the behavior of astrophysical and cosmological systems but also crucial tools for exploring complex phenomena such as black holes and wormholes.

In this work, we solve the $(n+2)$-dimensional EFE with non-zero cosmological constant $\Lambda$, 
\begin{equation}
\label{EFE}
    R_{A B} - \tfrac{R}{2} g_{A B} + \Lambda g_{A B} = 0\, ,
\end{equation}
for all $A, B \in \{ 1, \ldots, n + 2 \}$.
To do so, we use an algebraic flat subspaces method introduced in \cite{Sarmiento-Alvarado2023, Sarmiento-Alvarado2025}, which can be applied for $n > 1$.

In the present article, the set of $n \times n$ real matrices is denoted by $\mathbf{M}_n$, $I_n \in \mathbf{M}_n$ is the identity matrix, and $0$ is the $n \times n$ zero matrix.
For any matrix $A\in \mathbf{M}_n$, $A^T$, $\operatorname{tr} A$ and $\det A$ denotes its transpose, trace, and determinant, respectively.
$\mathbf{Sym}_m$ is the set of $m \times m$ symmetric matrices.
The special linear group of order $n$ is denoted by $SL ( n, \mathbb{R} )$, and its corresponding Lie algebra by $\mathfrak{sl} ( n, \mathbb{R} )$.
$\bar X$ represents the complex conjugate of $X$.
The $m$-dimensional Euclidean metric is denoted by $\hat{E}_m = (dx^1)^2 + \ldots (dx^m)^2$, the Minkowski metric by $\hat{\eta}_m = -dt^2 + \hat{E}_{m - 1}$, the $m$-sphere metric by $d\Omega_m^2$, and $\hat{T}_m = ( d\phi^1 )^2 + \ldots + ( d\phi^m )^2$ is the metric of $\mathbb{T}^m$, where $t ,x^k \in \mathbb{R}$ and $\phi^k \in [ 0, 2\pi )$ are angle variables, for all $k \in \{ 1, \ldots, m \}$.

The paper is structured as follows: the field equations are deduced in Section \ref{section: field eqs}. To solve them, we consider two cases: the metric components depending on one variable, discussed in Section \ref{section: one variable}, and on two variables, treated in Section \ref{section: two variables}. Section \ref{section: cosmic exp} studies a solution in the context of cosmology. Section \ref{section: conclusions} presents our Conclusions.

\section{Field equations}   %
\label{section: field eqs}

In this paper we consider an $(n+2)$-dimensional space with $n\geq 2$.
We assume that the space is endowed with a metric and contains $n$ commuting Killing vectors.
Then, we can choose a coordinate system where the metric depends only on two variables $x^1$ and $x^2$, so that the metric tensor is given as (see \cite{Matos89, 10.1063/1.529991})
\begin{equation}
\label{metric tensor}
    \hat g = f [ ( dx^1 )^2 + ( dx^2 )^2 ] + g_{\mu \nu} dx^\mu dx^\nu ,
\end{equation}
where the metric functions $f$ and $g_{\mu \nu}$ depend on $x^1$ and $x^2$, and $\mu, \nu \in \{ 3, \ldots, n + 2 \}$.

In order to write the Ricci tensor, let us define the matrix $g$ as
\begin{equation}
    ( g )_{\mu \nu} = g_{\mu \nu}\, .
\end{equation}
Now, we perform the change of variables as follows: $z = x^1 + i x^2$ and $\bar{z} = x^1 - i x^2$.
From \cite{Matos89, Sarmiento-Alvarado2023} we have that the non-zero components of Ricci tensor are given as
\begin{equation}
\begin{aligned}
    R_{1 1} & = P + Q_z + Q_{\bar z} ,
\\
    R_{2 2} & = P - Q_z - Q_{\bar z} ,
\\
    iR_{1 2} & =  Q_{\bar z} - Q_z ,
\\  \tensor{R}{_\mu ^\nu}
&   = - \frac{1}{f \rho } \  [
        \  ( \rho g_{, {\bar z}} g^{-1} )_{, z}
        + \  ( \rho g_{, z} g^{-1} )_{, {\bar z}}  ] ,
\end{aligned}
\end{equation}
where
\begin{align}
    P
&   = -2 (\ln f \rho )_{, {\bar z} z}
    - \frac{1}{2} \operatorname{tr} ( g_{, z} g^{-1} g_{, \bar{z}} g^{-1} ) ,
\\
    Q_{Z}
&   = ( \ln \rho )_{, Z} ( \ln f )_{, Z}
    - ( \ln \rho )_{, Z Z}
    - \frac{1}{4} \operatorname{tr} ( g_{,Z} g^{-1} )^2 ,
\end{align}
$\rho = \sqrt{ -\det g }$ and $Z \in \{ z, \bar z \}$.
In the remainder of this section, the index $Z$ continues to take these two values.

Eq. \eqref{EFE} can be rewritten as
\begin{equation}
    R_{A B} = \frac{2}{n} \Lambda g_{A B} .
\end{equation}
Then, the field equations in terms of $f$, $\rho$ and $g$ are as follows:
\begin{align}\label{field eq f}
    ( \ln f \rho )_{, Z}
&   = \frac{\rho_{, Z Z}}{\rho_{, Z}}
    + \frac{
        \tr ( g_{,Z} g^{-1} )^2
    }{4 ( \ln \rho )_{, Z}}
\\\label{integrability condition}
    (\ln f \rho )_{, z {\bar z}}
&   = - \frac{\Lambda}{n} f
    -\frac{1}{4} \tr ( g_{, z} g^{-1} g_{, \bar{z}} g^{-1} )
\end{align}
and
\begin{equation}
\label{inhomogeneous chiral eq}
    ( \rho g_{, {\bar z}} g^{-1} )_{, z} +( \rho g_{, z} g^{-1} )_{, {\bar z}} = - \frac{2 \Lambda}{n}f \rho I_n 
\end{equation}
Additionally, we obtain a field equation for $\rho$ from the trace of the inhomogeneous chiral equation \eqref{inhomogeneous chiral eq},
\begin{equation}
\label{field eq det}
    \rho_{, z \bar{z}} = - \frac{\Lambda}{2} f \rho \, .
\end{equation}

In what follows, we demonstrate that Eqs. \eqref{field eq f}, \eqref{inhomogeneous chiral eq} and \eqref{field eq det} imply Eq. \eqref{integrability condition}.
To achieve this, we differentiate Eq. \eqref{field eq f} with respect to $\bar{Z}$, obtaining
\begin{equation}
\label{2do derivative f}
    ( \ln f \rho )_{, Z \bar Z}
    = \left[ \frac{\rho_{, Z Z}}{\rho_{, Z}} \right]_{, \bar Z}
    + \frac{
        \left[ \operatorname{tr} ( g_{, Z} g^{-1} )^ 2 \right]_{, \bar Z}
    }{
        ( \ln \rho )_{, Z}
    } - \frac{
        ( \ln \rho )_{, Z \bar Z}
        \operatorname{tr} ( g_{, Z} g^{-1} )^ 2
    }{
        [ ( \ln \rho )_{, Z} ]^2
    } \, .
\end{equation}
In this section, the index $\bar Z$ gives the complex conjugate of the index $Z$.
Using Eq. \eqref{field eq det}, we find
\begin{equation}
\label{1st term field eq f}
    \left[ \frac{\rho_{, Z Z}}{\rho_{, Z}} \right]_{, \bar Z}
    = - \frac{\Lambda f}{2 ( \ln \rho )_{, Z}} \left[
        ( \ln f \rho )_{, Z}
        - \frac{\rho_{, Z Z}}{\rho_{, Z}}
    \right] .
\end{equation}
Replacing Eq. \eqref{field eq f} into Eq. \eqref{1st term field eq f}, we get
\begin{equation}
\label{1st term field eq f g}
    \left[ \frac{\rho_{, Z Z}}{\rho_{, Z}} \right]_{, \bar Z}
    = - \frac{\Lambda f}{8} \frac{
        \operatorname{tr} ( g_{, Z} g^{-1} )^ 2 
    }{
        [ ( \ln \rho )_{, Z}]^2
    } \, .
\end{equation}
Eq. \eqref{inhomogeneous chiral eq} can be rewritten as
\begin{equation}
    ( g_{, z} g^{-1} )_{, {\bar z}}
    + ( g_{, {\bar z}} g^{-1} )_{, z}
    + ( \ln \rho )_{\bar z} g_{, z} g^{-1}
    + ( \ln \rho )_z g_{,\bar z} g^{-1}
    = - \frac{2 \Lambda}{n} f I_n \, .
\end{equation}
Multiplying it by $g_{, Z} g^{-1}$ and then computing its trace, one obtains
\begin{equation}
\label{2do term field eq f}
    \frac{
        \left[ \operatorname{tr} ( g_{, Z} g^{-1} )^ 2 \right]_{, \bar Z}
    }{
        ( \ln \rho )_{, Z}
    }
    = - \frac{
        ( \ln \rho )_{\bar Z}
    }{
        ( \ln \rho )_{, Z}
    } \operatorname{tr} ( g_{, Z} g^{-1} )^2
    - \operatorname{tr} [ g_{, Z} g^{-1} g_{, {\bar Z}} g^{-1} ]
    - \frac{4 \Lambda}{n} f \, ,
\end{equation}
where we have used the identity
\begin{equation}
    \operatorname{tr} \left[  g_{, Z} g^{-1} ( g_{, {\bar Z}} g^{-1} )_{, Z} \right]
    = \frac{1}{2} \left[ \operatorname{tr} ( g_{, Z} g^{-1} )^ 2 \right]_{, \bar Z} .
\end{equation}
Finally, substituting Eqs. \eqref{1st term field eq f g} and \eqref{2do term field eq f} in Eq. \eqref{2do derivative f}, Eq. \eqref{integrability condition} is obtained.

In order to simplify the field equations, we perform the transformation
\begin{equation}
\label{transformation g}
    g \to -\rho ^{-2/n} g \, ,
\end{equation}
then $\det g = (-)^{n+1}$, so that $g$ is a matrix in $SL ( n, \mathbb{R} )$.
Consequently, the field equations \eqref{field eq f} and \eqref{inhomogeneous chiral eq} change to 
\begin{align}
\label{field eq f normalize}&
    ( \ln f \rho^{1 - 1/n} )_{, Z}
    = \frac{\rho_{, Z Z}}{\rho_{, Z}}
    + \frac{\operatorname{tr} ( g_{,Z} g^{-1} )^2}{4 ( \ln \rho )_{, Z}} \, ,
\\\label{chiral eq}&
    ( \rho g_{, {\bar z}} g^{-1} )_{, z} +( \rho g_{, z} g^{-1} )_{, {\bar z}} = 0 \, ,
\end{align}
respectively.
Now, the metric $g_{\mu \nu}$ is given by $g_{\mu \nu} = - \rho^{2/n} (g)_{\mu \nu}$.

In \cite{Sarmiento-Alvarado2023, Sarmiento-Alvarado2025}, we introduced the flat subspaces method and obtained solutions to the chiral equation \eqref{chiral eq} given by
\begin{equation}
\label{sol chiral eq}
    g ( z, \bar{z} ) = -\exp( \xi^a ( z, \bar{z} ) A_a ) g_0 \, ,
\end{equation}
where $\{ A_a, a=1,2,\ldots , m \} \subset \mathfrak{sl} ( n, \mathbb{R} )$ is a set of pairwise commuting constant matrices, $g_0$ is a constant matrix belonging to  $\mathcal{I} (\{ A_a, a=1,2,\ldots , m \}) = \{ M \in \mathbf{Sym}_n : A_a M = M A_a^T \text{ for all } a \}$ with $\det g_0 = -1$, and $\{ \xi^a ( z, \bar{z} ) \}$ is a set of solutions to the generalized Laplace equation
\begin{equation}
\label{gen Laplace eq}
    ( \rho \xi^a_{, z} )_{, \bar{z}} + ( \rho \xi^a_{, \bar{z}} )_{, z} = 0 \, .
\end{equation}
In \cite{Sarmiento-Alvarado2023}, $g ( z, \bar{z} )$ was computed considering only one matrix $A$, while in \cite{Sarmiento-Alvarado2025} a set $\{ A_a, a=1,2,\ldots , m \}$ of two or more matrices was considered.
It is important to note that only two or more matrices can be employed for $n > 2$.
Using the solution \eqref{sol chiral eq}, the field equation for $f$ becomes
\begin{equation}
\label{diff eq f}
    ( \ln f \rho^{1 - 1/n} )_{, Z}
    = \frac{\rho_{, Z Z}}{\rho_{, Z}}
    + \frac{\operatorname{tr} A_a A_b }{4 } \frac{ \xi^a_{, Z} \xi^b_{, Z} }{( \ln \rho )_{, Z}}
\end{equation}
and the metric $g_{\mu \nu}$ is then given by $g_{\mu \nu} = \rho^{2/n}  \exp( \xi^a ( z, \bar{z} ) A_a ) g_0$.
Here, the indices $a $ and $b$ run from 1 to $m$, where $m \in \{ 1, \ldots, n - 1 \}$ is fixed.

Observe that a constant matrix $g \in \mathbf{Sym}_n$ is also a solution to the chiral equation \eqref{chiral eq}.
This solution is obtained by setting $\xi^a = 0$ or $A_a = 0$ for all $a$ in Eq. \eqref{sol chiral eq}.
For the examples below, we consider $g$ and $g_0$ to be either the identity matrix $I_n$, or the matrix $\operatorname{diag} [ -1, I_{n - 1} ]$, or a block-diagonal matrix formed from these.

In summary, we only solve Eqs. \eqref{field eq det}, \eqref{gen Laplace eq} and \eqref{diff eq f} to obtain exact solutions to the EFE, which is discussed in the following two sections.

\section{One variable}  %
\label{section: one variable}

In this section, we assume that $\rho$, $f$ and $\xi^a$ depend on an arbitrary parameter $\lambda = \lambda ( z, \bar{z} )$.
Then Eqs.\eqref{field eq det}, \eqref{field eq f normalize} and \eqref{gen Laplace eq} are changed to
\begin{align}
\label{field eqs one parameter det}
    \rho_{, \lambda} \lambda_{, z \bar{z}}
    + \rho_{, \lambda \lambda} \lambda_{, z} \lambda_{, \bar{z}}
&   = - \frac{\Lambda}{2} f \rho \, ,
\\\label{field eqs one parameter f}
    \frac{\lambda_{, Z Z}}{( \lambda_{, Z} )^2}
    + \frac{\rho_{, \lambda \lambda}}{\rho_{, \lambda}}
    + \frac{\operatorname{tr} ( A_a A_b )}{4} \frac{ \xi^a_{, \lambda} \xi^b_{, \lambda} }{( \ln \rho )_{, \lambda}}
&   = ( \ln f \rho^{1 - 1/n} )_{, \lambda} \, ,
\\\label{field eqs one parameter chiral}
    \rho \xi^a_{, \lambda} \lambda_{, z \bar{z}}
    + ( \rho \xi^a_{, \lambda} )_{, \lambda} \lambda_{, z} \lambda_{, \bar{z}}
&   = 0 \, .
\end{align}

In order to simplify the above equations, we multiply Eq. \eqref{field eqs one parameter chiral} by $\rho \xi^b_{, \lambda}$, and then contract the result with $\operatorname{tr} ( A_a A_b )$, obtaining
\begin{equation}
\label{arbitrary parameter eq}
    \varphi \lambda_{, z \bar{z}}
    + \frac{1}{2} \varphi_{, \lambda} \lambda_{, z} \lambda_{, \bar{z}}
    = 0 \, ,
\end{equation}
where we have defined
\begin{equation}
\label{sqrt norm matrices A}
    \varphi (\lambda) = \operatorname{tr}( A_a A_b ) \rho^2 \xi^a_{, \lambda} \xi^b_{, \lambda} \, .
\end{equation}
To solve Eq. \eqref{arbitrary parameter eq}, we suppose that $\lambda$ depends on a parameter $\zeta = \zeta ( z, \bar{z} )$, which satisfies the Laplace equation
\begin{equation}
\label{Laplace eq harmonic parameter}
    \zeta_{, z \bar{z}} = 0 \, .
\end{equation}
Then, Eq. \eqref{arbitrary parameter eq} becomes $\varphi \lambda_{, \zeta \zeta} + \frac{1}{2} \varphi_{, \zeta} \lambda_{, \zeta} = 0$.
It follows that $\zeta_{, \lambda} = \sqrt{\vert \varphi \vert}$.
Substituting it into Eq. \eqref{sqrt norm matrices A}, we find $\ell = \operatorname{tr}( A_a A_b ) \rho^2 \xi^a_{, \zeta} \xi^b_{, \zeta} = \pm 1$.
This definition of $\ell$ allows to consider the case of constant $g$.
In this case, the constant parameter $\ell$ takes the value 0, and $\zeta$ is not subject to any constraints, that is, $\zeta$ can be any harmonic function.
Therefore, the field equations in terms of $\zeta$ are given as
\begin{align}
\label{field eqs harmonic parameter det}
    \rho_{, \zeta \zeta} \zeta_{, z} \zeta_{, \bar{z}}
&   = - \frac{\Lambda}{2} f \rho \, ,
\\\label{field eqs harmonic parameter f}
    \frac{\zeta_{, Z Z}}{( \zeta_{, Z} )^2}
    + \frac{\rho_{, \zeta \zeta}}{\rho_{, \zeta}}
    + \frac{\ell}{4 \rho \rho_{, \zeta}}
&   = ( \ln f \rho^{1 - 1/n} )_{, \zeta} \, ,
\\\label{field eqs harmonic parameter chiral}
    ( \rho \xi^a_{, \zeta} )_{, \zeta}
&   = 0 \, .
\end{align}

From Eq. \eqref{field eqs harmonic parameter chiral}, we have
\begin{equation}
    \rho \xi^a_{, \zeta} = \mathscr{C}^a \, ,
\end{equation}
where $\mathscr{C}^a \in \mathbb{R}$ are constants.
Rewriting $\ell$ in terms of these constants, one obtains
\begin{equation}
    \ell = \operatorname{tr}( A_a A_b ) \mathscr{C}^a \mathscr{C}^b \, .
\end{equation}
Recall that $\ell$ has only three values: -1, 0, and 1.
On the other hand, Eq. \eqref{field eqs harmonic parameter f} imply
\begin{equation}
\label{condition parameter}
    \frac{\zeta_{, z z}}{( \zeta_{, z} )^2}
    = \frac{\zeta_{, \bar{z} \bar{z}}}{( \zeta_{, \bar{z}} )^2} \, .
\end{equation}
The general solution of Eq. \eqref{Laplace eq harmonic parameter} is given as $\zeta ( z, \bar{z} ) = U (z) + V (z)$, where $U$ is a holomorphic function and $V$ is an antiholomorphic function.
Substituting this solution into Eq. \eqref{condition parameter}, we obtain
\begin{equation}
    \frac{U_{, z z}}{( U_{, z} )^2}
    = \frac{V_{, \bar{z} \bar{z}}}{( V_{, \bar{z}} )^2}
    = \mathscr{B} \, ,
\end{equation}
where $\mathscr{B}$ is a real constant.
Solving this, we find
\begin{equation}
\label{sols harmonic parameter}
\begin{array}{ll}
    \zeta = z + \bar{z}
&   \text{for } \mathscr{B} = 0 \, ,
\\  \zeta = \frac{-1}{\mathscr{B}} \ln ( z + C_z ) ( \bar{z} + C_{\bar{z}} )
&   \text{for } \mathscr{B} \neq 0 \, ,
\end{array}
\end{equation}
where $C_z, C_{\bar{z}} \in \mathbb{C}$ are constant.
Thus, Eq. \eqref{field eqs harmonic parameter det} changes to
\begin{equation}
\label{field eq det harmonic parameter}
\begin{array}{rl}
    \rho_{, \zeta \zeta} = - \frac{\Lambda}{2} f \rho
&   \text{for } \mathscr{B} = 0 \, ,
\\  \mathscr{B}^{-2} e^{\mathscr{B} \zeta} \rho_{, \zeta \zeta} = - \frac{\Lambda}{2} f \rho
&   \text{for } \mathscr{B} \neq 0\, .
\end{array}
\end{equation}
Substituting Eqs. \eqref{sols harmonic parameter} and \eqref{field eq det harmonic parameter} into Eq. \eqref{field eqs harmonic parameter f} leads to the following non-linear differential equation:
\begin{equation}
\label{nonlinear diff eq det}
    \frac{\rho_{, \zeta \zeta \zeta}}{\rho_{, \zeta \zeta}}
    = \frac{\rho_{, \zeta \zeta}}{\rho_{, \zeta}}
    + \frac{1}{n} \frac{\rho_{, \zeta}}{\rho}
    + \frac{\ell}{4 \rho \rho_{, \zeta}} \, .
\end{equation}

In order to solve Eq. \eqref{nonlinear diff eq det}, we define $\tilde{\gamma} = \rho_{, \zeta}$ and assume that $\tilde{\gamma} = \tilde{\gamma} ( \rho ( \zeta ) )$.
Using the chain rule, the equation can be written as
\begin{equation}
    \frac{ \rho \tilde{\gamma}_{, \rho \rho} }{ \tilde{\gamma}_{, \rho} }
    = \frac{1}{n}
    + \frac{\ell}{4 \tilde{\gamma}^2} \, .
\end{equation}
Then, the change of variable $\rho = e^{ \tilde{\rho} }$ gives
\begin{equation}
    \frac{ \tilde{\gamma}_{, \tilde{\rho} \tilde{\rho}} }{ \tilde{\gamma}_{, \tilde{\rho}} }
    = 1 + \frac{1}{n}
    + \frac{\ell}{4 \tilde{\gamma}^2} \, .
\end{equation}
Now, interchanging the independent and dependent variables, one obtains the following differential equation:
\begin{equation}
    -\frac{ \tilde{\rho}_{, \tilde{\gamma} \tilde{\gamma}} }{( \tilde{\rho}_{, \tilde{\gamma}} )^2}
    = 1 + \frac{1}{n}
    + \frac{\ell}{4 \tilde{\gamma}^2} \, .
\end{equation}
Integration with respect to $\tilde{\gamma}$ gives
\begin{equation}
    \tilde{\gamma}_{, \tilde{\rho}}
    = \frac{n + 1}{n} \tilde{\gamma}
    - \frac{\ell}{4 \tilde{\gamma}}
    + \mathscr{D} \, ,
\end{equation}
where $\mathscr{D} \in \mathbb{R}$ is an integration constant.
Again, let us perform a change of variables as $\gamma = \rho_{, \zeta} + \frac{n \mathscr{D}}{2 ( n + 1 )}$, which permits to write the above differential equation as follows:
\begin{equation}
\label{simplified det eq}
    \rho \gamma_{, \rho}
    = \frac{n + 1}{n} \frac{
        \gamma^2 - \frac{ n^2 \mathscr{E} }{ 4 ( n + 1 )^2 }
    }{
        \gamma - \frac{ n^ \mathscr{D} }{ 2 ( n + 1 ) }
    } \, ,
\end{equation}
where $\mathscr{E} = \mathscr{D}^2 + \frac{n + 1}{n} \ell$.
Solving this, we get 
\begin{equation}
\label{det gamma}
\begin{array}{llll}
\left (\frac{\rho}{\rho_0} \right)^{1 + 1/n}
    = \gamma
    \exp \left( \frac{p_0}{\gamma} \right) ,
&   0 < \vert \gamma \vert
&   \text{if } \mathscr{E} = 0
&   \multirow{3}{*}{for $\Lambda < 0$ ,}
\\  \left (\frac{\rho}{\rho_0} \right)^{1 + 1/n}
    = \sqrt{ \gamma^2 - \gamma_+^2 }
    \exp \left(
        p_+ \coth^{-1} \frac{\gamma}{\gamma_+}
    \right) ,
&   \gamma_+ < \vert \gamma \vert
&   \text{if } \mathscr{E} > 0
&
\\  \left (\frac{\rho}{\rho_0} \right)^{1 + 1/n}
    = \sqrt{ \gamma^2 + \gamma^2_- }
    \exp \left( -p_- \tan^{-1} \frac{\gamma}{\gamma_-} \right) ,
&   \gamma \in \mathbb{R}
&   \text{if } \mathscr{E} < 0
&
\\  \left (\frac{\rho}{\rho_0} \right)^{1 + 1/n}
    = \sqrt{ \gamma_+^2 - \gamma^2 } \exp \left(
        p_+ \tanh^{-1} \frac{\gamma}{\gamma_+}
    \right) ,
&   \vert \gamma \vert < \gamma_+ 
&   \text{if } \mathscr{E} > 0
&   \text{for } \Lambda > 0 \, ,
\end{array}
\end{equation}
where $\rho_0$ is a positive integration constant, $\gamma_\pm = \dfrac{n \sqrt{\pm \mathscr{E}}}{2 (n + 1)}$, $p_\pm = \dfrac{\mathscr{D}}{\sqrt{\pm \mathscr{E}}}$ and $p_0 = \dfrac{n \mathscr{D}}{2 (n + 1)}$.

For a non-constant matrix $g$, we need to determine the parameters $\xi^a$.
Applying the chain rule to the differential equation $\rho \xi^a_{, \zeta} = \mathscr{C}^a$, we obtain the following:
\begin{equation}
    \xi^a_{, \gamma}
    = \dfrac{
        \dfrac{n \mathscr{C}^a}{n + 1}
    }{
        \gamma^2
        - \dfrac{n^2 \mathscr{E}}{4 ( n + 1 )^2}
    } \, .
\end{equation}
Integration gives the following result:
\begin{equation}
\begin{array}{lll}
    \xi^a =
    \xi^a_0
    -\dfrac{ 2 q_0 }{ \gamma }&   \text{if } \mathscr{E} = 0
&   \multirow{3}{*}{for $\Lambda < 0$ ,}
\\  \xi^a =
    \xi^a_0
    -2 q^a_+
    \coth^{-1} \dfrac{ \gamma }{ \gamma_+ }
&   \text{if } \mathscr{E} > 0
&
\\  \xi^a =
    \xi^a_0
    + 2 q^a_-
    \tan^{-1} \dfrac{ \gamma }{ \gamma_- }
&   \text{if } \mathscr{E} < 0
&
\\  \xi^a =
    \xi^a_0
    - 2 q^a_+ \tanh^{-1} \frac{ \gamma }{ \gamma_+ }
&   \text{if } \mathscr{E} > 0
&   \text{for } \Lambda > 0 \, ,
\end{array}
\end{equation}
where $\xi^a_0 \in \mathbb{R}$ are integration constants, $q^a_\pm = \dfrac{\mathscr{C}^a}{ \sqrt{\pm \mathscr{E}} }$ and $q^a_0 = \dfrac{n \mathscr{C}^a}{2 (n + 1)}$.

Once we know $p_\pm$ and $q_\pm$, we can relate them.
Let us rewrite the definition of $\mathscr{E}$ as $\pm 1 = (\tfrac{\mathscr{D}}{\sqrt{\pm \mathscr{E}}})^2 + \tfrac{n + 1}{n} \tfrac{\ell}{\pm \mathscr{E}}$.
Expressing $\ell$ in terms of $q_\pm$, we get $\tfrac{\ell}{\pm \mathscr{E}} = \operatorname{tr} ( A_a A_b ) q_\pm^a q_\pm^b$.
Therefore, $\pm 1 = p_\pm^2 + \tfrac{n + 1}{n} \operatorname{tr} ( A_a A_b ) q_\pm^a q_\pm^b$.
For $\mathscr{E} = 0$, we have $0 = \mathscr{D}^2 + \frac{n + 1}{n} \ell$.
Multiplying this by $4 \tfrac{( n + 1 )^2}{n^2}$ and then using the definition of $p_0$ and $q_0$, we obtain $p_0^2 + \tfrac{n + 1}{n} \operatorname{tr} ( A_a A_b ) q_0^a q_0^b = 0$.

When $\mathscr{B} = 0$, $\gamma$ is related to $x^1$ by $\tfrac{d \gamma}{d x^1} = - \Lambda f \rho$.
We can use $\gamma$ as a coordinate instead of $x^1$.
Then, the metric tensor reads
\begin{equation}
\label{metric tensor gamma}
    \hat g
    = \dfrac{d \gamma^2}{ \Lambda^2 f \rho^2 }
    + f ( d x^2 )^2
    + \rho^{2/n} g_{\mu \nu} d x^\mu d x^\nu \, ,
\end{equation}
where $g_{\mu \nu} = \exp( \xi^a A_a ) g_0$.
Using Eq. \eqref{simplified det eq}, $f$ can be expressed in terms of $\gamma$ as
\begin{equation}
\label{sol f}
    f = \frac{2 ( n + 1 ) }{n \Lambda \rho^2} \left(
        \frac{n^2 \mathscr{E}}{4 ( n + 1 )^2}
        -  \gamma^2
    \right) .
\end{equation}
Therefore,
\begin{equation}
\label{sol metric 1 var}
    \hat{g}
    = \frac{n}{ 2 ( n + 1 ) \vert \Lambda \vert } \frac{d\gamma^2}{\mathfrak{a}}
    + \sqrt[n + 1]{\mathfrak{a}} \left(
        \frac{ 2 ( n + 1 ) }{ n \vert \Lambda \vert \rho_0^2 } \frac{( dx^2 )^2}{\mathfrak{b}^{2 n}}
        + \rho_0^\frac{2}{n} \mathfrak{b}^2 g_{\mu \nu} dx^\mu dx^\nu
    \right) \, ,
\end{equation}
where
\begin{equation}
\begin{array}{lllll}
    \mathfrak{a}
    = \gamma^2 ,
&   \mathfrak{b}
    = \exp( \tfrac{p_0}{n + 1} \tfrac{1}{\gamma} ) ,
&   \text{if } \mathscr{E} = 0
&   \multirow{3}{*}{for $\Lambda < 0$,}
\\  \mathfrak{a}
    = \gamma^2 - \gamma_+^2 ,
&   \mathfrak{b}
    = \exp( \tfrac{p_+}{n + 1} \coth^{-1} \tfrac{\gamma}{\gamma_+} ) ,
&   \text{if } \mathscr{E} > 0
&
\\  \mathfrak{a}
    = \gamma^2 + \gamma_-^2 ,
&   \mathfrak{b}
    = \exp( \tfrac{- p_-}{n + 1} \tan^{-1} \frac{\gamma}{\gamma_-} ) ,
&   \text{if } \mathscr{E} < 0
&
\\  \mathfrak{a}
    =  \gamma_+^2 - \gamma^2 ,
&   \mathfrak{b}
    = \exp( \tfrac{p_+}{n + 1} \tanh^{-1} \frac{\gamma}{\gamma_+} ) ,
&   \text{if } \mathscr{E} > 0
&   \text{for } \Lambda > 0 \, .
\end{array}
\end{equation}

\begin{example}
\label{example: AdS dS}
    The Anti-de Sitter and the de Sitter metrics.
\end{example}
Let $g = \operatorname{diag} [ -1, I_{n - 1} ]$, and assume that $\Lambda < 0$ and $\mathscr{E} = 0$.
Since $g$ is a constant matrix, $\ell = 0$, so that $p_0 = 0$, hence $\mathfrak{b} = 0$.
Performing the change of variables as $\sqrt[n + 1]{\gamma} = \tfrac{n \rho_0}{2 x}$, we obtain the $(n+2)$-dimensional Anti-de Sitter ($AdS_{n + 2}$) metric of radius $R = \sqrt{\tfrac{ n ( n + 1 )}{-2 \Lambda}}$ in horospheric coordinates (see \cite{10.1007/3-540-46671-1_5}):
\begin{equation}
\label{AdS metric}
    \hat{g} = R^2 \frac{
        dx^2 + \hat{\eta}_{n + 1}
    }{x^2} \, ,
\end{equation}
where we have set $\rho_0 = (\tfrac{2 R}{n})^\frac{n}{n + 1}$.

In static coordinates, the $AdS_{n + 2}$ metric \eqref{AdS metric} reads
\begin{equation}
    \hat{g}
    = -( 1 + \tfrac{r^2}{R^2} ) dt^2
    + ( 1 + \tfrac{r^2}{R^2} )^{-1} dr^2
    + r^2 d\Omega_n^2 \, .
\end{equation}
Now, we perform a Wick rotation $R = iR_s$, to obtain
\begin{equation}
    \hat{g}
    = -( 1 - \tfrac{r^2}{R_s^2} ) dt^2
    + ( 1 - \tfrac{r^2}{R_s^2} )^{-1} dr^2
    + r^2 d\Omega_n^2 \, ,
\end{equation}
which is the $(n + 2)$-dimensional de Sitter ($dS_{n + 2}$) metric of radius $R_s = \sqrt{\tfrac{ n ( n + 1 )}{2 \Lambda}}$ in static coordinates.
Note that $\Lambda$ is now positive.

\begin{example}
\label{example: Birmingham metric}
    The Birmingham metric \cite{Danny_Birmingham_1999}.
\end{example}
We set $g = \operatorname{diag} [ -1, I_{n - 1} ]$, $\Lambda < 0$, $\mathscr{E} > 0$ and $p_+ = 1$.
Expressing the cotangent function in terms of the natural logarithm function, we have $\mathfrak{b} = \sqrt[2 ( n + 1 )]{\tfrac{\gamma + \gamma_+}{\gamma - \gamma_+}}$.
Now, we make the change of variable given as $r = \tfrac{n + 1}{\vert \Lambda \vert \rho_0} \sqrt[n + 1]{ \gamma + \gamma_+ }$ and $\phi^k = \tfrac{n \rho_0^{1 + 1/n}}{2 R^2} x^k$, where $R = \sqrt{\tfrac{n ( n + 1 )}{2 \vert \Lambda \vert}}$ and $\phi^k \in [ 0, 2 \pi )$ for all $k \in \{ 4, \ldots, n + 2 \}$.
In addition, we perform two Wick rotations: $x^2 = it$ and $\tfrac{n \rho_0^{1 + 1/n}}{2 R^2} x^3 = i\phi^3$, with $\phi^3 \in [ 0, 2 \pi )$.
Therefore,
\begin{equation}
    \hat{g}
    = -( \tfrac{r^2}{R^2} - \tfrac{2 m}{r^{n - 1}} ) dt^2
    + \frac{dr^2}{ \tfrac{r^2}{R^2} - \tfrac{2 m}{r^{n - 1}} }    
    + r^2 \hat{T}_n \, ,
\end{equation}
where $m = \tfrac{2 \gamma_+}{n \rho_0} (\tfrac{n + 1}{\vert \Lambda \vert \rho_0})^n$.

\section{Two variables} %
\label{section: two variables}

Substituting Eq. \eqref{field eq det} into Eq. \eqref{diff eq f} we get
\begin{equation}
\label{system diff eqs nD}
\begin{array}{l}
    \dfrac{ \rho_{, z z \bar{z} } }{ \rho_{, z \bar{z} } }
    = \dfrac{ \rho_{, z z } }{ \rho_{, z} }
    + \dfrac{1}{n} \dfrac{ \rho_{, z } }{ \rho }
    + \dfrac{\operatorname{tr} A_a A_b}{4} \dfrac{\xi^a_{, z} \xi^b_{, z}}{( \ln \rho )_{, z} } \  ,
\\  \dfrac{ \rho_{, \bar{z} \bar{z} z } }{ \rho_{, \bar{z} z } }
    = \dfrac{ \rho_{, \bar{z} \bar{z} } }{ \rho_{, \bar{z}} }
    + \dfrac{1}{n} \dfrac{ \rho_{, \bar{z} } }{ \rho }
    + \dfrac{\operatorname{tr} A_a A_b}{4} \dfrac{ \xi^a_{, \bar{z}} \xi^b_{, \bar{z}} }{( \ln \rho )_{, \bar{z}} } \  .
\end{array}
\end{equation}
To solve the above system of differential equations, we propose that $\rho ( z, \bar{z} ) = X (z)  Y (\bar{z})$.
Then
\begin{equation}
\label{field eqs red nD}
    \operatorname{tr} ( A_a A_b ) \frac{\xi^a_{, z}}{( \ln X )_{, z}} \frac{\xi^b_{, z}}{( \ln X )_{, z}}
    = \operatorname{tr} ( A_a A_b ) \frac{\xi^a_{, \bar{z} }}{( \ln Y )_{, \bar{z} }} \frac{\xi^b_{, \bar{z} }}{( \ln Y )_{, \bar{z} }}
    = -\frac{4}{n} \, .
\end{equation}

Now, Eq. \eqref{gen Laplace eq} can be rewritten as
\begin{equation}
\label{gen Laplace eq XY nD}
    \frac{
        \xi^a_{, z}
    }{
       ( \ln X )_{, z}
    }
    + \frac{
        \xi^a_{, \bar{z}}
    }{
       ( \ln Y )_{, \bar{z}}
    }
    = -\frac{ 2 \xi^a_{, z \bar{z} } }{
        ( \ln X )_{, z} ( \ln Y )_{, \bar{z}}
    } \, .
\end{equation}
Assuming that the parameters $\xi^a$ satisfy
\begin{equation}
\label{laplace eq parameters}
    \xi^a_{, z \bar{z} } = 0 \, ,  
\end{equation}
one obtains
\begin{equation}
\label{sol parameter}
    \xi^a ( z, \bar{z} ) = U^a ( z ) + V^a ( \bar{z} ) \, ,
\end{equation}
where $U^a$ are holomorphic functions and $V^a$ are antiholomorphic functions.
Given that $\xi^a$ are real, then $\operatorname{Im} ( \xi^a ) = 0$ implies $\operatorname{Im} (V^a) = - \operatorname{Im} (U^a)$.

Replacing Eqs. \eqref{laplace eq parameters} and \eqref{sol parameter} into Eq. \eqref{gen Laplace eq XY nD}, we find
\begin{equation}
\label{UV const}
    \frac{ U^a_{, z} }{ ( \ln X )_{, z} }
    = -\frac{ V^a_{, \bar{z}} }{ ( \ln Y )_{, \bar{z}} }
    = -i\mathscr{C}^a \, ,
\end{equation} 
where $\mathscr{C}^a \in \mathbb{R}$ are constants.
It is convenient to define $\mathscr{C}_a = \operatorname{tr} ( A_a A_b ) \mathscr{C}^b$.
Substituting Eq. \eqref{UV const} into \eqref{field eqs red nD}, we obtain
\begin{equation}
\label{consts eq}
    \mathscr{C}^a \mathscr{C}_a = \frac{4}{n} \, .
\end{equation}
Contracting Eq. \eqref{UV const} with $\mathscr{C}_a$, and then solving for $X$ and $Y$, we get
\begin{equation}
\label{XY functions}
    X = X_0 \exp\left( i \tfrac{n}{4} \mathscr{C}_a U^a \right)
    \text{ and }
    Y = Y_0 \exp\left( -i \tfrac{n}{4} \mathscr{C}_b V^b \right) ,
\end{equation}
so that $\rho = \rho_0 \exp\left( i \tfrac{n}{4} \mathscr{C}_a ( U^a - V^a ) \right)$, where $\rho_0 = X_0 Y_0$ and $X_0, Y_0 \in \mathbb{C}$ are constant.
For $\rho$ to be real, we set $X_0 = Y_0 = 1$ and assume that $\operatorname{Re} ( V^b )= \operatorname{Re} ( U^b )$.

The metric component $f$ is given by $f = -\frac{2}{\Lambda} ( \ln X )_{, z} ( \ln Y )_{, \bar{z}}$.
Using Eq. \eqref{XY functions}, we write $f$ as $f = S_{, z \bar{z}}$,
where
\begin{equation}
    S = -\frac{n^2}{8 \Lambda} \mathscr{C}_a \mathscr{C}_b U^a \bar{U}^b \, .
\end{equation}
Once we know $\xi^a$ and $\rho$, we can compute the metric component $g_{\mu \nu}$ as follows:
\begin{equation}
    g_{\mu \nu}
    = \exp( -\mathscr{C}_a \operatorname{Im}( U^a ) I_n + 2\operatorname{Re} (U^a) A_a ) g_0 \, .
\end{equation}
Finally, the metric \eqref{metric tensor} can be expressed in terms of $z$ and $\bar{z}$:
\begin{equation}
    \hat{g}
    = S_{, z \bar{z}} dz d\bar{z}
    + g_{\mu \nu} dx^\mu dx^\nu \, .
\end{equation}

If we consider one matrix $A$, $S$ reduces to $S = - \tfrac{n^2}{8 \Lambda} ( \mathscr{C}^1 \operatorname{tr} A^2 )^2 U \bar{U}$.
From Eq. \eqref{consts eq}, we have $( \mathscr{C}^1 )^2 \operatorname{tr} A^2 = \tfrac{4}{n}$.
Substituting it into $S$, we get $S = - \tfrac{2}{\Lambda ( \mathscr{C}^1 )^2}  U \bar{U}$.
We set $- \tfrac{2}{\Lambda ( \mathscr{C}^1 )^2} = 1$, so $S = U \bar{U}$ and $\mathscr{C}^1 = \pm \sqrt{\tfrac{2}{\vert \Lambda \vert}}$.
Using the known value of $\mathscr{C}^1$, one can compute
\begin{equation}
\label{one dim trace A2}
    \operatorname{tr} A^2 = \frac{2}{n} \vert \Lambda \vert
\end{equation}
and thus $\mathscr{C}_1 = \pm \tfrac{2}{n} \sqrt{ 2 \vert \Lambda \vert}$.
Finally, we perform the following change of coordinates: $X^1 = \operatorname{Re} (U)$ and $X^2 = \operatorname{Im} (U)$.
Therefore,
\begin{equation}
\label{one dim metric}
    \hat g = ( dX^1 )^2 + ( dX^2 )^2
    + g_{\mu \nu} dx^\mu dx^\nu \, ,
\end{equation}
 where $g_{\mu \nu}$ is given by
\begin{equation}
\label{int metric one dim A}
    g_{\mu \nu} = \exp(
        \pm \tfrac{2}{n} \sqrt{2 \vert \Lambda \vert} X^2 I_n
        + 2 X^1 A
    ) g_0 \, .
\end{equation}

\begin{example}\label{example: two subspaces}
    Let $\sigma$ be a positive constant and $A = \sigma \operatorname{diag} [ I_\frac{n}{2}, -I_\frac{n}{2} ]$ a matrix in $\mathfrak{sl} ( n, \mathbb{R} )$, with even $n$.
\end{example}
From Eq. \eqref{one dim trace A2}, we have $\sigma = \tfrac{\sqrt{ 2 \vert \Lambda \vert }}{n}$. Computing $g_{\mu \nu}$ given by Eq. \eqref{int metric one dim A} results in
\begin{equation}
    g_{\mu \nu} = \operatorname{diag} \left[
        \exp( \tfrac{4}{n} \sqrt{\vert \Lambda \vert} ( \pm X^2 + X^1 ) ) I_\frac{n}{2} ,
        \exp( -\tfrac{4}{n} \sqrt{\vert \Lambda \vert} ( \pm X^2 - X^1 ) ) I_\frac{n}{2}
    \right] .
\end{equation}
The constant matrix $g_0$ has the form $g_0 = \operatorname{diag} [ M_1, M_2 ] \in \mathbf{Sym}_n$, where $M_1, M_2 \in \mathbf{Sym}_\frac{n}{2}$ are constant.
Since $\det g_0 = -1$, we have $\det M_1 \det M_2 = -1$.
To determinate the matrices $M_1$ and $M_2$, we assume that $\hat{g}$ tends to the Minkowski metric as $\Lambda$ approaches 0.
Hence, $M_1 = \operatorname{diag} [ -1, I_{\frac{n}{2} - 1} ]$ and $M_2 = I_\frac{n}{2}$.
Now, making the change of variables $\sqrt{2} x^1 = \pm X^2 + X^1 $ and $\sqrt{2} x^2 = \pm X^2 - X^1$, leads to
\begin{equation}
    \hat{g}
    = ( dx^1 )^2 + \exp( 2 \tfrac{x^1}{R} ) \hat{\eta}_\frac{n}{2}
    + ( dx^2 )^2 + \exp( 2 \tfrac{x^2}{R} ) \hat{E}_\frac{n}{2} \, ,
\end{equation}
where $R = \tfrac{n}{ 2 \sqrt{-\Lambda} }$.
Finally, performing the change of variables as $y^k = R \exp( -\tfrac{x^k}{R} )$ for all $k \in \{ 1, 2 \}$, results in
\begin{equation}
\label{sol AdSH}
    \hat{g}
    = R^2 \frac{
        ( dy^1 )^2 + \hat{\eta}_\frac{n}{2}
    }{( y^1 )^2}
    + R^2 \frac{
        ( dy^2 )^2 + \hat{E}_\frac{n}{2}
    }{( y^2 )^2} \, .
\end{equation}
The second term is the metric of the hyperbolic upper half-space $H^{\frac{n}{2} + 1}$ \cite{Voight2021}.
This solution is the direct topological product $AdS_{ \frac{n}{2} + 1 } \times H^{\frac{n}{2} + 1}$. 
When $n = 2$, the metric \eqref{sol AdSH} reduces to the Anti‑Nariai solution.

In order to obtain solutions with positive cosmological constant, we write the $AdS_{ \frac{n}{2} + 1 }$ and the $H^{\frac{n}{2} + 1}$ metrics in static coordinates.
Then
\begin{equation}
    \hat{g}
    = -( 1 + \tfrac{\rho^2}{R^2} ) dt^2
    + ( 1 + \tfrac{\rho^2}{R^2} )^{-1} d\rho^2
    + \rho^2 d\Omega^2_{\frac{n}{2} - 1}
    + ( 1 + \tfrac{r^2}{R^2} ) d\chi^2
    + ( 1 + \tfrac{r^2}{R^2} )^{-1} dr^2
    + r^2 d\Omega^2_{\frac{n}{2} - 1} \, .
\end{equation}
Finally, we make the change of variable as $\chi = R \phi$, with $\phi \in [ 0, 2 \pi )$, and a Wick rotation $R \to i \tilde{R}$, where $\tilde{R} = \tfrac{n}{ 2 \sqrt{\Lambda} }$, $\Lambda > 0$.
Hence,
\begin{equation}
\label{sol dS sphere}
    \hat{g}
    = -( 1 - \tfrac{\rho^2}{\tilde{R}^2} ) dt^2
    + ( 1 - \tfrac{\rho^2}{\tilde{R}^2} )^{-1} d\rho^2
    + \rho^2 d\Omega^2_{\frac{n}{2} - 1}
    + ( 1 - \tfrac{r^2}{\tilde{R}^2} ) \tilde{R}^2 d\phi^2
    + ( 1 - \tfrac{r^2}{\tilde{R}^2} )^{-1} dr^2
    + r^2 d\Omega^2_{\frac{n}{2} - 1} \, .
\end{equation}
The first three terms in Eq. \eqref{sol dS sphere} form the $dS_{\frac{n}{2} + 1}$ metric, while the last three terms form the metric of a $(\tfrac{n}{2} + 1)$-sphere of radius $\tilde{R}$.
In Appendix \ref{appendix: static coordinates}, we introduce the static coordinates for the hyperbolic and the $m$-sphere metrics.
The solution \eqref{sol dS sphere} is the direct topological product $dS_{ \frac{n}{2} + 1 } \times S^{\frac{n}{2} + 1}$.
In the case $n = 2$, the metric \eqref{sol dS sphere} is known as the Nariai solution.

\begin{example}\label{example: matrix S0}
    Let $A = \sigma \operatorname{diag} [ I_{n - 1}, 1 - n ] \in \mathfrak{sl} ( n, \mathbb{R} )$, with positive $\sigma$.
\end{example}
Given that $\operatorname{tr} A^2 = ( n - 1 ) n \sigma^2$, by Eq. \eqref{one dim trace A2}, we have $\sigma = \tfrac{1}{n} \sqrt{\tfrac{2 \vert \Lambda \vert}{n - 1}}$.
Substituting the known value of $\sigma$ into Eq. \eqref{int metric one dim A}, and then making the change of variables as $\sqrt{n} x^1 = \pm X^2 + \tfrac{X^1}{\sqrt{n - 1}}$ and $\sqrt{n} x^2 = \pm X^2 - \sqrt{n - 1} X^1$, we obtain
\begin{equation}
    g_{\mu \nu}
    = \operatorname{diag} \left[
        \exp( 2 \sqrt{2 n \vert \Lambda \vert} x^1 ) I_{n - 1},
        \exp( 2 \sqrt{2 n \vert \Lambda \vert} x^2 )
    \right] g_0\, .
\end{equation}
If the matrix $g_0$ belongs to $\mathcal{I} (A)$ then it has the form $g_0 = \operatorname{diag} [ M_0, \tfrac{-1}{\det M_0} ]$, where $M_0 \in \mathbf{Sym}_{n - 1}$ is constant.
Note that $M_0 = I_{n - 1}$ and $M_0 = \operatorname{diag} [ -1, I_{n - 2} ]$ satisfy $\hat{g} \to \hat{\eta}_{n + 2}$ as $\Lambda \to 0$.
Each choice yields a different exact solution.
For both solutions, we perform the change of variables as follows: $y^k = R_k \exp( -\sqrt{ 2 n \vert \Lambda \vert } x^k )$ for all $k \in \{ 1, 2 \}$, where $R_1 = \sqrt{ n - 1} R_2$ and $R_2 = \tfrac{1}{ n \sqrt{-\Lambda} }$.
This leads to
\begin{align}\label{sol AdS2H}
    \hat{g}
&   = R_1^2 \frac{
        ( dy^1 )^2 + \hat{E}_{n - 1}
    }{( y^1 )^2}
    + R_2^2 \frac{
        ( dy^2 )^2 - dt^2
    }{( y^2 )^2} \, ,
\\\label{sol H2Ads}
    \hat{g}
&   = R_1^2 \frac{
        ( dy^1 )^2 + \hat{\eta}_{n - 1}
    }{( y^1 )^2}
    + R_2^2 \frac{
        ( dy^2 )^2 + dx^2
    }{( y^2 )^2} \, .
\end{align}
The solution \eqref{sol AdS2H} is the direct topological product of $AdS_2 \times H^n$, while the solution \eqref{sol H2Ads} is the direct topological product of $AdS_n \times H^2$.
From these solutions, we obtain the Anti-Nariai solution for $n = 2$.

In what follows, we obtain solutions with a positive cosmological constant.
Similarly to Example \ref{example: two subspaces}, we write the metrics \eqref{sol AdS2H} and \eqref{sol H2Ads} in terms of static coordinates,
\begin{align}
    \hat{g}
&   = ( 1 + \tfrac{r^2}{R_1^2} ) d\chi^2
    + ( 1 + \tfrac{r^2}{R^1} )^{-1} dr^2
    + r^2 d\Omega^2_{n - 2}
    - ( 1 + \tfrac{\rho^2}{R_2^2} ) dt^2
    + ( 1 + \tfrac{\rho^2}{R_2^2} )^{-1} d\rho^2 \, ,
\\
    \hat{g}
&   = -( 1 + \tfrac{\rho^2}{R_1^2} ) dt^2
    + ( 1 + \tfrac{\rho^2}{R_1^2} )^{-1} d\rho^2
    + \rho^2 d\Omega^2_{n - 2}
    + ( 1 + \tfrac{r^2}{R_2^2} ) d\chi^2
    + ( 1 + \tfrac{r^2}{R_2^2} )^{-1} dr^2 \, .
\end{align}
Finally, we perform two Wick rotations: $R_1 \to i \tilde{R}_1$ and $R_2 \to i \tilde{R}_2$, and the following change of variables: $\chi = R \phi$, where $\tilde{R}_1 = \sqrt{ n - 1} \tilde{R}_2$, $\tilde{R}_2 = \tfrac{1}{ n \sqrt{\Lambda} }$, $\Lambda > 0$ and $\phi \in [ 0, 2 \pi )$.
Thus,
\begin{align}\label{sol dS2 sphere}
    \hat{g}
&   = - ( 1 - \tfrac{\rho^2}{\tilde{R}_2^2} ) dt^2
    + ( 1 - \tfrac{\rho^2}{\tilde{R}_2^2} )^{-1} d\rho^2
    + ( 1 - \tfrac{r^2}{\tilde{R}_1^2} ) \tilde{R}_1^2 d\phi^2
    + ( 1 - \tfrac{r^2}{\tilde{R}^1} )^{-1} dr^2
    + r^2 d\Omega^2_{n - 2} \, ,
\\\label{sol 2-sphere dS}
    \hat{g}
&   = -( 1 - \tfrac{\rho^2}{\tilde{R}_1^2} ) dt^2
    + ( 1 - \tfrac{\rho^2}{\tilde{R}_1^2} )^{-1} d\rho^2
    + \rho^2 d\Omega^2_{n - 2}
    + ( 1 - \tfrac{r^2}{\tilde{R}_2^2} ) \tilde{R}^2 d\phi^2
    + ( 1 - \tfrac{r^2}{\tilde{R}_2^2} )^{-1} dr^2 \, .
\end{align}
Both solutions are direct topological products of $dS_2 \times S^n$ and $dS_n \times S^2$, respectively, and reduce to the Nariai solution for $n = 2$.
The metric \eqref{sol dS2 sphere} is obtained in \cite{PhysRevD.70.024002}.

\section{Cosmic expansion}    
\label{section: cosmic exp}

In this section, we will obtain a solution whose metric components depend on a variable. The solution is then studied it in the context of cosmology.

We set $n = 2$, $\Lambda < 0$ and $\mathscr{E} > 0$ in the metric \eqref{sol metric 1 var}.
Let $A = \operatorname{diag} [ \sigma, -\sigma ] \in \mathfrak{sl} ( 2, \mathbb{R} )$, where $\sigma \in \mathbb{R}$ is a positive constant.
Since $g_0$ is a constant matrix belonging to $\mathcal{I} (A)$, it has the form $g_0 = \operatorname{diag} [ c_0, -c_0^{-1} ]$, where $c_0$ is a positive constant.
Performing the changes of variables as $\gamma = \gamma_+ \cosh \sqrt{3 \vert \Lambda \vert} \chi$, the metric components $\mathfrak{a}$, $\mathfrak{b}$ and $\xi$ are then given by $\mathfrak{a} = \gamma_+^2 \sinh^2 \sqrt{3 \vert \Lambda \vert} \chi$, $\mathfrak{b} = \coth^\frac{p_+}{3} \tfrac{\sqrt{3 \vert \Lambda \vert}}{2} \chi$, and $\xi = \xi_0 - 2 q_+ \ln \coth \tfrac{\sqrt{3 \vert \Lambda \vert}}{2} \chi$, respectively.
Thus, $g_{\mu \nu} = \operatorname{diag} [ c_0 e^{\sigma \xi_0} \tanh^{2 \sigma q_+} \tfrac{\sqrt{3 \vert \Lambda \vert}}{2} \chi, -c_0^{-1} e^{-\sigma \xi_0} \coth^{2 \sigma q_+} \tfrac{\sqrt{3 \vert \Lambda \vert}}{2} \chi ]$.

It is convenient to define $p = -p_+$ and $q = \sigma q_+$, then $p$ and $q$ satisfy $1 = p^2 + 3 q^2$.
Additionally, we set $\rho_0 = \sqrt{-\tfrac{3}{\Lambda}}$ and make the following change of variables:  $z = x^2$, $x = \rho_0 c_0 e^{\sigma \xi_0} x^3$.
Finally, we perform three Wick rotations: $\chi = i t$, $y = i \tfrac{\rho_0}{c_0 e^{\sigma \xi_0}} x^4$, and
$\sqrt{-3 \Lambda}i \to \sqrt{3 \Lambda}$, which allows to obtain
\begin{equation}
\label{cosmological metric}
    \hat{g}
    = -dt^2
    + \gamma_+^\frac{2}{3} \sinh^\frac{2}{3} \sqrt{3 \Lambda} t \left(
        \tanh^{ \frac{2}{3} p + 2 q} \frac{\sqrt{3 \Lambda} t}{2} dx^2
        + \tanh^{ \frac{2}{3} p - 2 q} \frac{\sqrt{3 \Lambda} t}{2} dy^2
        + \coth^\frac{4 p}{3} \frac{\sqrt{3 \Lambda} t}{2} dz^2
    \right) .
\end{equation}
Here, $x, y, z$ are the Cartesian coordinates, and $t$ is the time.
Now, the cosmological constant is positive.
This metric is the Bianchi I.
Since $q$ cannot be zero, this metric describes a homogeneous and anisotropic universe.

However, we can compute an asymptotic expansion for this metric.
To determine it, we rewrite the definition of the hyperbolic tangent function as $\tanh u= 1 - \tfrac{2 e^{-2 u}}{1 + e^{-2 u}}$.
Using the fact that $\tfrac{2 e^{-2 u}}{1 + e^{-2 u}} < 1$ for $u > 1$, one gets $\tanh^s u = 1 - 2 s e^{-2 u} + \ldots$, where $s \in \mathbb{R}$ is a constant.
Expressing the powers of the hyperbolic sine function as $\sinh^s u = \tfrac{e^{ s u }}{2^s} - \tfrac{s}{2^s} e^{( s -2 ) u} + \ldots$, leads to
\begin{equation}
    \hat{g}
    = -dt^2
    + \gamma_+^\frac{2}{3} \sinh^\frac{2}{3} \sqrt{3 \Lambda} t \left( dx^2 + dy^2 + dz^2 \right)
    - \tfrac{4}{3} (\tfrac{\gamma_+}{2})^\frac{2}{3} \tfrac{
        ( p + 3 q ) dx^2
        + ( p - 3 q ) dy^2 
        - 2 p dz^2
    }{\exp(\sqrt{\frac{\Lambda}{3}} t)}
    + \ldots
\end{equation}
For large times, the contribution of the exponential function can be neglected, which means that initially the universe is anisotropic and then becomes isotropic.

We can assume a scale factor for large times given by $a = \sqrt[3]{ \gamma_+ \sinh \sqrt{3 \Lambda} t }$.
So, the Hubble and deceleration parameters are given as
\begin{align}
\label{H anisotropic}
    H
&   = \frac{\dot a}{a}
    = \sqrt{\frac{\Lambda}{3}} \coth \sqrt{3 \Lambda} t \, ,
\\\label{q anisotropic}
    q
&   = -\frac{\ddot a}{a H^2}= 2 - 3 \tanh^2 \sqrt{3 \Lambda} t \, ,
\end{align}
respectively.
We set $a ( t_0 ) = 1$ by convention.
From Eq. \eqref{H anisotropic}, we find the present time $t_0 = \frac{1}{\sqrt{3 \Lambda}} \tanh^{-1} \sqrt{\frac{\Lambda}{3 H_0^2}}$.
Using it, we determine $\gamma_+ = \sqrt{ \tfrac{3 H_0^2}{\Lambda} - 1 }$.
Evaluating Eq. \eqref{q anisotropic} at $t = t_0$, we get $q_0 = q ( t_0 ) = 2 - \frac{\Lambda}{H_0^2}$.
Thus, the cosmological constant can be expressed as $\Lambda = ( 2 - q_0 ) H_0^2$.

If the metric \eqref{cosmological metric} is expressed as $\hat{g} = -dt^2 + a^2 (g)_{\mu \nu} dx^\mu dx^\nu$, the EFE can be rewritten as
\begin{align}
    3 \dot H
    + \tfrac{1}{4} \operatorname{tr} ( \dot g g^{-1} )^2
    + 3 H^2
&   = \Lambda
\\\label{field eq g}  
    \tfrac{1}{2 a^3}( a^3 \dot g g^{-1} )_{, t}
    + ( \dot H + 3 H^2 ) I_3
    & = \Lambda I_3 \, ,
\end{align}
where $g = \operatorname{diag} [ \tanh^{ \frac{2}{3} p + 2 q} \frac{\sqrt{3 \Lambda} t}{2}, \tanh^{ \frac{2}{3} p - 2 q} \frac{\sqrt{3 \Lambda} t}{2}, \coth^\frac{4 p}{3} \frac{\sqrt{3 \Lambda} t}{2} ]$ and $\dot f$ denotes the derivative of $f$ with respect to the time.
Differentiating $g$ with respect to the time, we get $a^3 \dot g g^{-1} = 2 \gamma_+ \sqrt{\tfrac{\Lambda}{3}} \operatorname{diag} [ p + 3 q, p - 3 q, -2 p ]$.
Then, $\operatorname{tr} ( \dot g g^{-1} )^2 = \tfrac{8 \Lambda \gamma_+^2}{a^6}$ and $\dot H + 3 H^2 = \Lambda$, so that $H^2 = \tfrac{\Lambda}{3} + \tfrac{\Lambda \gamma_+^2}{3 a^6}$.
Let $\Omega_\Lambda = \tfrac{\Lambda}{3 H^2}$ and $\Omega_s = \tfrac{\Lambda \gamma_+^2}{3 a^6 H^2}$ be the density parameters.
Evaluating at $t = t_0$, $\Omega_{\Lambda, 0} = \Omega_\Lambda (t_0) = \tfrac{\Lambda}{3 H_0^2}$ and $\Omega_{s, 0} = \Omega_s (t_0) = \tfrac{\Lambda \gamma_+^2}{3 H_0^2}$ are obtained.
Now, let us rewrite the Friedmann equation as
\begin{equation}
\label{friedmann eq}
    H = H_0 \sqrt{
        \Omega_{\Lambda, 0}
        + \frac{\Omega_{s, 0}}{a^6}
    } \, .
\end{equation}

In order to express $\gamma_+$ in terms of $\Omega_{\Lambda, 0}$ and $\Omega_{s, 0}$, we write $\Omega_{s, 0} = \Omega_{\Lambda, 0} \gamma_+^2$ and then substitute it into Eq. \eqref{friedmann eq} evaluated at $t = t_0$, obtaining $\gamma_+ = \sqrt{\tfrac{\Omega_{s, 0}}{\Omega_{\Lambda, 0}}}$.
Therefore, we write the scale factor as
\begin{equation}
\label{scale factor}
    a = \sqrt[6]{\tfrac{\Omega_{s, 0}}{\Omega_{\Lambda, 0}}} \sqrt[3]{\sinh 3 \sqrt{\Omega_{\Lambda, 0}} H_0 t} \, .
\end{equation}
Comparing Eq. \eqref{scale factor} with Eq. (56) of \cite{PhysRevD.92.103004} for an isotropic spacetime with a positive cosmological constant and stiff matter, we see that the scale factors are equal.

In what follows, we show that $\operatorname{tr} ( \dot g g^{-1} )^2$ satisfies an equation of state similar to that of stiff matter.
For this, let us rewrite Eq. \eqref{field eq g} as $( \dot g g^{-1} )_{, t} + 3 H \dot g g^{-1} + 2 ( \dot H + 3 H^2 - \Lambda ) I_3 = 0$.
Multiplying by $\dot g g^{-1}$ and then computing its trace, we get $( \operatorname{tr} ( \dot g g^{-1} )^2 )_{, t} + 6 H \operatorname{tr} ( \dot g g^{-1} )^2 = 0$.
This means that anisotropy behaves as a perfect fluid with the equation of state $\omega = 1$.

\begin{figure}
    \centering
    \begin{subfigure}{0.45\textwidth}
        \includegraphics[width=1\linewidth]{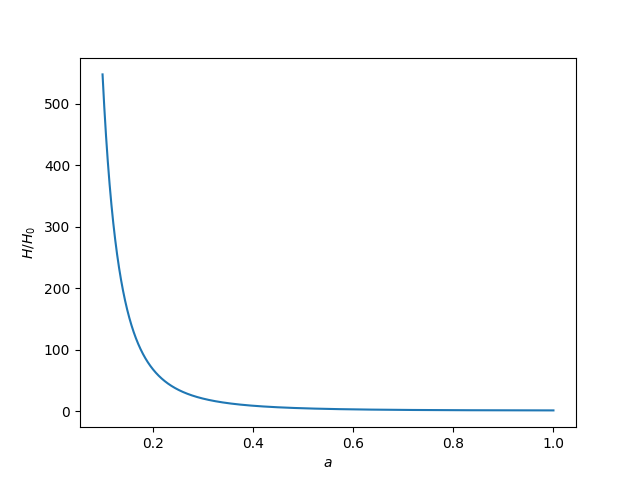}
        \caption{$H$.}
        \label{fig: H}
    \end{subfigure}
\hfill
    \begin{subfigure}{0.45\textwidth}
        \includegraphics[width=1\linewidth]{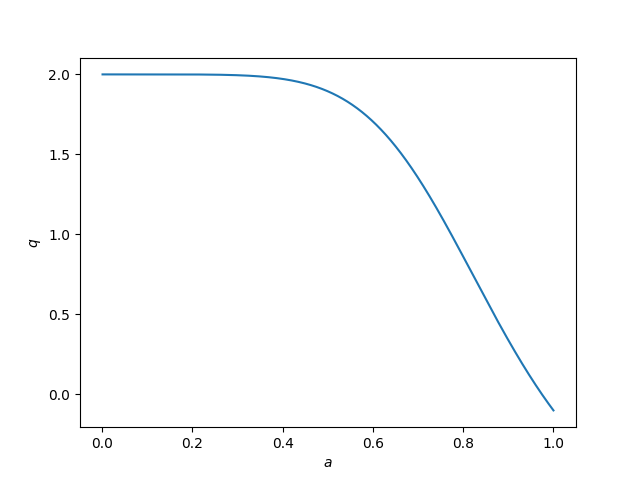}
        \caption{$q$.}
        \label{fig: q}
    \end{subfigure}
\caption{Behavior of $H$ and $q$ as function of $a$.}
\label{fig: cosmological parameters}
\end{figure}

Figure \ref{fig: cosmological parameters} shows the behavior of the cosmological parameters $H$ and $q$ as functions of $a$, where we set $\Omega_{\Lambda, 0} = 0.7$.
\begin{equation}
    q = 2
    - \frac{ 3 \Omega_{\Lambda, 0} }{
        \Omega_{\Lambda, 0}
        + \frac{\Omega_{s, 0}}{a^6}
    } \, .
\end{equation}
The figure illustrates that $H$ has a vertical asymptote at $a = 0$ and is a strictly decreasing function that takes the value $H_0$ at $a = 1$. The parameter $q$ is also a decreasing function, but it slowly decreases until $a \approx 0.5$, then rapidly decreases and crosses the $a$-axis at $a = \sqrt[6]{2 \tfrac{\Omega_{s, 0}}{\Omega_{\Lambda, 0}}} \approx 0.97464$.

\section{Conclusions}   %
\label{section: conclusions}

In this work, we obtained exact solutions to the EFE in higher dimensions with non-zero cosmological constant using the flat subspaces method.
These solutions depend on a subset $\{ A_a, a=1,2,\ldots, m\} \subset \mathfrak{sl} ( n, \mathbb{R} )$ of pairwise commuting constant matrices and on a constant matrix $g_0 \in \mathcal{I} (\{ A_a \})$.
This set and matrix were determined for $n > 1$ in the previous work \cite{Sarmiento-Alvarado2023, Sarmiento-Alvarado2025} using tools from linear algebra and the properties of the group $SL ( n, \mathbb{R} )$.

We divided our solutions into two parts.
In the first part, we assume that the metric components depend on one variable, while in the second part they depend on two variables.
The solutions of the first part depend on a constant parameter $\mathscr{B}$.
When $\mathscr{B} = 0$, four types of solutions are obtained: three for $\Lambda < 0$ and one for $\Lambda > 0$.
From them, we obtained the de Sitter metric, the Anti-de Sitter metric, and the Birmingham metric in higher dimensions, see our Examples \ref{example: AdS dS} and \ref{example: Birmingham metric}.
The present work does not consider solutions with $\mathscr{B} \neq 0$.

In the case where the metric components depend on two variables, the method used only gives solutions with a negative cosmological constant.
However, as we have seen in Examples \ref{example: two subspaces} and \ref{example: matrix S0}, we can get solutions with a positive cosmological constant by means of Wick rotations in an adequate system of coordinates.
The solutions obtained in Examples \ref{example: two subspaces} and \ref{example: matrix S0} are direct topological products of $AdS_{\frac{n}{2} + 1} \times H^{\frac{n}{2} + 1}$, $dS_{\frac{n}{2} + 1} \times S^{\frac{n}{2} + 1}$, $AdS_n \times H^2$, $dS_n \times S^2$, $AdS_2 \times H^n$ and $dS_2 \times S^n$, which generalizes the Nariai and the Anti-Nariai solutions to higher dimensions.

In Section \ref{section: cosmic exp}, we studied a solution obtained under Wick rotations in the context of cosmology.
This solution describes a homogeneous and anisotropic expanding spacetime.
This spacetime becomes isotropic for large values of time, and after some time, it begins to expand at an accelerating rate.
If we write this solution as $\hat{g} = -dt^2 + a^2 g_{\mu \nu} dx^\mu dx^\nu$, where the metric components $g_{\mu \nu}$ depend on $t$ and $\det g_{\mu \nu} = 1$, then the EFE reduce to the Friedmann equation with dark energy and a term similar to stiff matter.
This term is the contribution of $g_{\mu \nu}$, which is not constant and is a product of the anisotropy of spacetime.

We showed how to obtain known solutions through a change of variables and Wick rotations.
However, there are still solutions that have not been studied, because each set $\{ A_a, a=1,2,\ldots, m\}$ gives a different solution to the EFE.

\appendix
\section*{Appendix: Static coordinates}   %
\label{appendix: static coordinates}

\subsection*{Hyperbolic space}

In this section, we will express the hyperbolic metric,
\begin{equation}
    \hat{H}_{m + 1} = R^2 \frac{ du^2 + \hat{E}_m }{u^2} ,\, u \in ( 0, \infty ) \, ,
\end{equation}
in terms of static coordinates.
To achieve this, we begin by making the following change of variables: $u = \exp(\tfrac{v}{R})$, with $v \in \mathbb{R}$, and expressing the Euclidean metric $\hat{E}_m$ in spherical coordinates.
Then,
\begin{equation}
    \hat{H}_{m + 1}
    = dv^2
    + R^2 \exp( -2 \tfrac{v}{R} ) ( dr^2 + r^2 d\Omega_{m -1}^2 ) \, .
\end{equation}
Here, $r \in [ 0, \infty )$ is the radial coordinate.
Again, we perform a change of variables: $\rho = R r \exp( -\tfrac{v}{R} )$, which gives
\begin{equation}
    \hat{H}_{m + 1}
    = ( 1 + \tfrac{\rho^2}{R^2} ) dv^2
    + 2 \tfrac{\rho}{R} dv d\rho
    + d\rho^2
    + \rho^2 d\Omega_{m -1}^2  \, .
\end{equation}
Finally, the change of variables $w = v + \tfrac{R}{2} \ln ( 1 + \tfrac{\rho^2}{R^2} )$ allows to obtain the result as follows:
\begin{equation}
    \hat{H}_{m + 1}
    = ( 1 + \tfrac{\rho^2}{R^2} ) dw^2
    + ( 1 + \tfrac{\rho^2}{R^2} )^{-1} d\rho^2
    + \rho^2 d\Omega_{m -1}^2 \, .
\end{equation}

\subsection*{$m$-sphere}

In this section, the $m$-sphere metric is determined in terms of static coordinates.
For this, we use the $m$-sphere equation of radius $R$,
\begin{equation}
\label{sphere eq}
     ( x^0 )^2 + ( x^1 )^2 + \ldots + ( x^m )^2 = R^2 \, .
\end{equation}
Defining $r = \sqrt{ ( x^2 )^2 + \ldots + ( x^m )^2 }$, Eq. \eqref{sphere eq} reduces to $( x^0 )^2 + ( x^1 )^2 = R^2 - r^2$, and the Euclidean metric can be rewritten in spherical coordinates as follows:
\begin{equation}
\label{euclidean metric spherical coord}
     \hat{E}_{m + 1} = ( dx^0 )^2 + ( dx^1 )^2 + dr^2 + r^2 d\Omega_{m - 2}^2 .
\end{equation}
Here, $r$ is the radial coordinate.
Now let us make the following changes of variables: $x^0 = \sqrt{ R^2 - r^2 } \cos \phi$ and $x^1 = \sqrt{ R^2 - r^2 } \sin \phi$, with $r \in [ 0, R ]$ and $\phi \in [ 0, 2 \pi )$, into the metric \eqref{euclidean metric spherical coord}. Then we obtain
\begin{equation}
    d\Omega_m^2
    = ( 1 - \tfrac{r^2}{R^2} ) R^2 d\phi^2
    + ( 1 - \tfrac{r^2}{R^2} )^{-1} dr^2
    + r^2 d\Omega_{m - 2}^2 \, .
\end{equation}

\printbibliography

\end{document}